\documentclass{amsart}
\usepackage{graphicx}
\theoremstyle{definition}

\theoremstyle{remark}

\numberwithin{equation}{section}

\begin{document}

\title[]{Density effects in forward scattering of resonant light in rubidium vapor}%
\author{Szymon Pustelny, Wojciech Lewoczko, Wojciech Gawlik}
\address{M. Smoluchowski Physical Institute, Jagiellonian University\\
Reymonta 4, 30-059 Krak\'{o}w, Poland}%
\email{pustelny@netmail.if.uj.edu.pl}%

\begin{abstract}
Transmission of light through an atomic sample placed between
crossed polarizers in a magnetic field is jointly determined by
Faraday rotation and net absorption: transmission increases with
rotation and decreases with absorption. Both rotation and
absorption are proportional to the atomic density $N_0$, hence, in
a certain range of $N_0$, the two effects may compete yielding a
distinct density dependence of the transmitted light. We have
studied such dependence in rubidium vapor for $N_0\approx
6.0\times 10^9\div 3.0\times 10^{11}$ at/cm$^3$ with resonant
laser light. We present interpretation of the competition effect
and discuss its possible application for atomic density
determination.
\end{abstract}
\maketitle
\section{Introduction}
Studies of magneto-optical effects in atoms and molecules have a
long tradition. The advent of tunable lasers stimulated renewed
interest in the nonlinear magneto-optics. Numerous papers are
describing various aspects of this nonlinearity (see
reference\cite{RMP} for recent review).  In the systems with
ground state (or other long-lived lower levels) of angular
momentum $J\neq0$, the magneto-optical nonlinearities are due to
optical pumping and Zeeman coherences. Optical pumping changes the
saturation conditions for a given transition by affecting the
sublevel populations whereas the Zeeman coherence allows
interference of quantum paths linking different sublevels. These
effects give rise to an optical anisotropy which can be easily
studied experimentally.

One important application of magneto-optical studies is the
determination of optical density and thus a measurement of
oscillator strength (dipole moment) of a given transition when the
number density is known or, inversely, of atomic densities when
the transition probability is known. Examples of such applications
of linear Faraday effect are the measurements of relative
oscillator strengths in neon \cite{Seka} and sodium \cite{WG-RFE}
and of atomic vapor density of optically thick potassium vapors
\cite{Happer}.

The present work describes experimental and theoretical analysis
of transmission of resonance laser light by atomic vapors
positioned between crossed polarizers in a weak longitudinal
magnetic field (forward scattering - FS) under conditions of
nonlinear light-matter interaction. In general, the FS signal,
i.e. the transmitted light intensity vs. magnetic field, depends
on the polarization anisotropy both in atomic absorption and
dispersion. In such context FS has been analyzed for conventional
light sources in early works of G. W. Series and coworkers
\cite{Corney}, and for laser light by W. Gawlik et al.\cite{GKNT}.
For a medium of length $L$, placed between crossed polarizers and
characterized by refractive indices $n_{\pm}$ and absorption
coefficients $\kappa_{\pm}$ for the $\sigma^{\pm}$ polarized
light, the intensity of the forward-scattered monochromatic light
of wavenumber $k$ is given by\cite{WG-RFE}
\begin{equation}I_{FS}=\frac{I_0}{4}[e^{-{\kappa_+k
L}}-e^{-{\kappa_-k
L}}]^2+I_0\sin^2[\frac{n_+-n_-}{2}kL]e^{-(\kappa_++\kappa_-)kL},\label{FS}
\end{equation}
where $I_0$ is the incident light intensity. The FS signal is
non-zero only if the medium possesses some optical anisotropy
between the $\sigma^{+}$ and $\sigma^{-}$ light polarizations. An
important cause of such anisotropy is a longitudinal magnetic
field which splits the $n_{\pm}(\omega)$ and
$\kappa_{\pm}(\omega)$ profiles. The first part of Eq. (\ref{FS})
describes the effect of medium dichroism whereas the second part
is due to its birefringence weighted by a net absorption. For the
light frequency $\omega$ tuned exactly to the atomic transition
frequency $\omega_{0}$ and for a linear splitting of atomic energy
levels due to magnetic field, the appearance of a nonzero FS
signal is caused exclusively by the second part of expression
(\ref{FS}), i.e. by the birefringence. The exact shape of the FS
signal dependence on the magnetic field is affected by a combined
effect of birefringence and total absorption. In the case of
nonlinear interaction with the light beam, this dependence may
have the form of a narrow sub-Doppler, or even subnatural,
resonances (see, e.g. references \cite{RMP,GKNT}). The two
properties of the medium, birefringence and total absorption,
scale differently with the atomic density which results in
specific density/temperature characteristics of the FS signals
discussed below. We hope that these characteristic density
dependences would be useful for determination of the optical
density of atomic samples.

The detailed mechanism of nonlinear FS signals has been
theoretically analyzed in several papers (see, e.g.
references\cite{Giraud-Cotton,Chen,Stahlberg,Schuller,Holmes,Lobodzinski})
at different levels of accuracy. Below, we present a
nonperturbative treatment that for low J values yields analytical
formulae.

In the next section we describe our setup and present experimental
results. In Sec. 3 we present theoretical calculations and
interpretation of the experiment. Finally, we conclude in Sec. 4.

\section{Experiment}

\subsection{Apparatus}

The experimental setup is shown in Fig. 1. We used an external
cavity diode laser tuned to the hfs component $F_g=3\rightarrow
F_e=2$ of the $^{85}$Rb $D_1$ line. The laser frequency was
stabilized by an external stabilization system based on
Doppler-free dichroism\cite{Stabilizacja}. The light beam of 2 mm
diameter was linearly polarized and its power was varied by
rotation of a $\lambda/2$-waveplate in the range 0 to 4 mW. The
laser light was passing through a 5 cm long cylindrical glass cell
of 2 cm diameter filled with the natural mixture of\ $^{85}$Rb and
$^{87}$Rb and no buffer gas. The cell was heated by a nonmagnetic
heater to temperatures between 20 $^o$C and
 60 $^o$C which yielded saturated vapors with densities in the
 range $6.0\times 10^9\div 3.0\times 10^{11}$ at/cm$^3$. A pair of
Helmholtz coils produced the longitudinal magnetic field of -3 G
to 3 G value. The cell and the coil system were surrounded by a
three-layer magnetic shield made of mu-metal. As assessed from the
resonance line-width, the shielding efficiency was better than
10$^4$. Polarizer P and crossed analyzer A were mounted outside
the shield\cite{Corney}. In that geometry, detector D placed
behind the analyzer registered only the FS signal due to rotation
of the polarization plane of the transmitted light and/or its
ellipticity. Part of the incident beam was directed to the
saturation spectroscopy system which served as frequency
reference.

\subsection{Results}

The FS signals, $I_{FS}(B)$, were recorded as a function of the
magnetic field $B$ for various temperatures $T$ and light
intensities $I_0$. The narrowest features in signals are caused by
nonlinear Faraday effect due to ground-state optical pumping and
Zeeman coherences. A typical plot of that part of the FS signal is
presented in Fig. 2a. The amplitude of the signal, defined as
$A_{FS}$ in Fig. 2a, depends on the light intensity and
concentration of the atomic vapors. In Fig. 3 we summarize the
results of the amplitude vs. temperature $T$ dependences taken for
various light intensities $I_0$.

For the studied range of intensities between 1 mW/cm$^2$ and 6
mW/cm$^2$, the signals exhibit characteristic temperature
dependence: for low temperature their amplitudes increase with $T$
then reach their maxima and subsequently decrease with $T$. The
temperature of the maximum amplitude shifts with the light
intensity. For high temperatures almost no light is transmitted
through the cell between crossed polarizers. For very high light
intensity (above 100 mW/cm$^2$), the maximum shifts even beyond
the range of temperatures available with our setup (Fig. 4).

We have verified that the amplitude of the FS signal is a
nonlinear function of the light intensity. This could be expected
as the intensities we were using were close to, or higher than,
the saturation intensity which for rubidium is 1.6 mW/cm$^2$. It
can be thus expected that the interplay between rotation and
absorption is affected by different saturation behavior of the
dispersive and absorptive atomic properties. Such different
saturation effects are manifest in Fig. 3 by a shift of the maxima
towards higher temperatures with increasing $I_0$.

\section{Interpretation}

\newcounter{popnr}\setcounter{popnr}{\value{equation}}
\addtocounter{popnr}{1}
\newcommand{\alpheqn}{\setcounter{equation}{0}
\renewcommand{\theequation}{\arabic{popnr}\alph{equation}}}
\newcommand{\reseteqn}{\setcounter{equation}{\value{popnr}}
\addtocounter{popnr}{1}
\renewcommand{\theequation}{\arabic{popnr}}}

The properties of the maxima in the $A_{FS}(T)$ dependences,
described above, are independent of the detailed mechanisms
responsible for the FS signals. The maximum occurs for linear, as
well as for nonlinear Faraday rotations. For nonlinear Faraday
effect, it appears no matter if the FS signals have natural or
sub-natural widths, i.e. if they are due to population
redistribution or Zeeman coherences. The detailed nature of atomic
birefringence and absorption mechanism affects only the range of
the concentrations (temperatures) where the maximum is visible.

This property can be illustrated by the following simple
arguments. For exact resonance, $\delta=0$, only the second part
of Eq.(\ref {FS}) differs from zero, hence
\begin{equation} \frac{I_{FS}}{I_0} = \sin^2 (\alpha N_0 )\exp(-\beta N_0), \label{alpha}
\end{equation}
\noindent where $\alpha/\beta=1/2(n_+ - n_-)/(\kappa_+ +
\kappa_-)$ and $N_0$ is the atomic concentration. For a given
atomic transition, cell length, magnetic field and light
intensities, $\alpha$ and $\beta$ are constants mutually related
by the Kramers-Kronig relation \cite{KK}. $I_{FS}/I_0$ is thus an
universal function of $N_0$, parameterized by the ratio
$\alpha/\beta$ but \emph{independent of the detailed form} of
$\alpha$ and $\beta$. Generally, the specific value of the ratio
depends on particular dependences of $\alpha$ and $\beta$ on
atomic energy level structure, mechanism of interaction with the
external fields and their intensities. However, for the value of
the magnetic field corresponding to maximum rotation $A_{FS}$ the
splitting of atomic magnetic sublevels is nearly equal to their
width which results in $\alpha/\beta$ of the order of 1. By virtue
of the Kramers-Kronig relation, the value of $\alpha/\beta$ at a
given $B$ is nearly insensitive to the details of the atomic
structure and to the nature of optical anisotropy. With
$\alpha/\beta=1$, expression (\ref{alpha}) represents a strongly
damped oscillatory dependence on $N_0$, dominated by the first
oscillation. These oscillations are nonlinear counterparts of
those described in reference \cite{WG-RFE}. Note, that the
appearance of the characteristic maximum of the $A_{FS}$
dependence on $N_0$ is possible only with the FS methodology and
cannot be observed by measuring only the rotation.

For more quantitative interpretation of our experimental results
we use semiclassical approach of forward scattering and density
matrix formalism\cite{Boyd} to calculate the refractive indices
$n$ and absorption coefficients $\kappa$
\reseteqn\alpheqn\begin{equation} n-1=\frac{\Re(\sum_i
d_i\langle\rho_{coh}^{(i)}\rangle_v)}{E\epsilon_0},
\label{refraction}\end{equation}
\begin{equation} \kappa=\frac{\Im(\sum_i d_i\langle\rho_{coh}^{(i)}
\rangle_v)}{E\epsilon_0}, \label{absorption}
\end{equation}
\noindent where $\langle\rho_{coh}^{(i)}\rangle_v$ are the
velocity averaged optical coherences for the i-th transition of a
given polarization, $\Re$ and $\Im$ denote their real and
imaginary parts, respectively, $d_i$ is the dipole moment of the
particular transition, $E$ the amplitude of the light electric
field, and $\epsilon_0$ is the electric permittivity of free
space.

In this paper we consider the simple model of atomic structure
$J_g=1/2\rightarrow J_e=1/2$ with two sublevels in the ground
state and two sublevels in the excited state the, so called, X
model. The $\sigma^+$ component of the propagating beam excites
atoms from the ground-state sublevel $|a\rangle$ ($m=-1/2$) to the
excited-state sublevel $|B\rangle$ ($m=1/2$). Analogically, the
$\sigma^-$ component causes transition from $|b\rangle$ ($m=1/2$)
to $|A\rangle$ ($m=-1/2$). The great advantage of this simple
model is its analytical solvability. Despite its simplicity, the X
model gives good qualitative agreement with the experimental
results. Furthermore it reproduces well the observed dependence of
the FS amplitude on the properties of the atomic medium, in
particular on the dipole moment and concentration of atoms in the
ground sublevels in thermal equilibrium.

To check to what extent the density dependence of $A_{FS}(N_0)$ is
insensitive to particular model, and to verify that the X model
does not oversimplify the interpretation, we have additionally
applied another model using the $\Lambda$-type atomic structure
(Fig. 5b). The second model allows proper analysis of the
coherence effects, such as coherent population trapping
(CPT)~\cite{Arim}, and their consequences for magneto-optical
properties of the medium. However, analytical solutions and
velocity integration are much more difficult within the second
model. For this reason, in this paper we concentrate exclusively
on the X model. Nevertheless, we have  verified by numerical
calculation with the $\Lambda$ model that Zeeman coherence effects
do not change qualitative predictions of the X model concerning
the temperature dependence of $A_FS(N)$. This is due to the fact
that the $I_FS(B)$ signals simulated by both models are
qualitatively very similar to the experimental ones, as seen in
Fig.2, so the $\alpha$ and $\beta$ quantities in Eq. (2) remain
very close, no matter what detailed model is being used.

Within our model, the refractive indices and absorption
coefficients for both circular polarizations are calculated using
equations (\ref{refraction}) and (\ref{absorption}) as \reseteqn
\alpheqn
\begin{equation}
n_+-1=\frac{d\Re(\langle\rho_{aB}\rangle_v)}{E\epsilon_0},\
n_--1=\frac{d\Re(\langle\rho_{bA}\rangle_v)}{E\epsilon_0},
\label{refforX}\end{equation}
\begin{equation}
\kappa_+=\frac{d\Im(\langle\rho_{aB}\rangle_v)}{E\epsilon_0},\
\kappa_-=\frac{d\Im(\langle\rho_{bA}\rangle_v)}{E\epsilon_0}\;.
\label{absforX}\end{equation} The semiclassical Hamiltonian of the
system is given by
\begin{equation}\reseteqn H=\hbar\omega_0(\rho_{AA}+\rho_{BB})+\frac{\hbar}{2}g_g\omega_B(\rho_{aa}-\rho_{bb})+
\frac{\hbar}{2}g_e\omega_B(\rho_{AA}-\rho_{BB})+dE(\rho_{aB}+\rho_{bA}+h.c.),\end{equation}
\noindent where $\rho_{ii}$ denotes population of the i-th state,
$g_g$, $g_e$ are the Land\'{e} factors of the ground and the
excited states, respectively, $\omega_0$ is the transition
frequency in zero magnetic field, and $\omega_B$ is the Larmor
frequency $\omega_B=\mu_BB/\hbar$ with $\mu_B$ being Bohr
magneton.

Evolution of the density matrix is described by the master
equation\reseteqn
\begin{equation}
\dot\rho=-\frac{i}{\hbar}[H,\rho]-\tilde\Gamma\rho\label{master},\end{equation}
where $\tilde\Gamma\rho$ stands for the relaxation operator.

Formula (\ref{master}) yields the following equations describing
evolution of the optical coherences in the X model \reseteqn
\alpheqn
\begin{equation}\dot{\rho}_{aB}=i(\omega_0+\omega_B'+i\Gamma)\rho_{aB}+i\beta(\rho_{aa}-\rho_{BB}),
\label{cohaB}\end{equation}
\begin{equation}\dot{\rho}_{bA}=i(\omega_0-\omega_B'+i\Gamma)\rho_{bA}+i\beta(\rho_{bb}-\rho_{AA}),
\label{cohaB}\end{equation} \noindent where
$\omega_B'=\frac{1}{2}(g_g+g_e)\,\omega_B$ is the effective
magnetic shift of the resonance frequency, $\beta=Ed / \hbar$ is
the Rabi frequency and $\Gamma$ is the relaxation constant of the
optical coherences.

Further calculations are performed in the rotating-wave and steady
state approximations. Under such conditions the optical coherences
become \reseteqn \alpheqn
\begin{equation}\tilde\rho_{aB}=-\frac{\beta}{\delta+\omega_B'+i\Gamma}
(\rho_{aa}-\rho_{BB}),\label{koherencja_aB}\end{equation}
\begin{equation}\tilde\rho_{bA}=-\frac{\beta}{\delta-\omega_B'+i\Gamma}
(\rho_{bb}-\rho_{AA}),\label{koherencja_bA}\end{equation}
\noindent where $\tilde\rho_{aB}$ and $\tilde\rho_{bA}$ are
slowly-varying coherences in the rotating-wave approximation and
$\delta=\omega_0-\omega$. Although only $\tilde\rho_{aB}$ is
calculated below, the analogical calculation can be performed for
$\tilde\rho_{Ba}$.

Relation (\ref{koherencja_aB}) couples the optical coherence with
atomic populations. Using Eq. (\ref {master}) to calculate atomic
populations and inserting them to (\ref{koherencja_aB}) one can
write \reseteqn
\begin{equation}\tilde\rho_{aB}=\beta n_{eq}(\delta+\omega_B'-i\Gamma)\frac{M_A(\delta)+2rG}{M_A(\delta)M_B(\delta)
+rG(M_A(\delta)+M_B(\delta))+2rG^2},\label{przed_rozkladem}\end{equation}
\noindent where $n_{eq}$ is the thermal equilibrium atomic
population of the ground state and other symbols are defined as
$$ M_{A}(\delta)=(\delta - \omega_B')^2+\Gamma^2
,$$ $$ M_{B}(\delta)=(\delta + \omega_B')^2+\Gamma^2 ,$$
$$ r=\frac{\gamma_e}{3\gamma_g}, $$
$$ G=\frac{2\beta^2\Gamma}{\gamma_e}, $$
\noindent with $\gamma_g$ and $\gamma_e$ being the relaxation
constants of the ground and excited states, respectively. Relation
(\ref{przed_rozkladem}) is an analitical expression valid for any
light intensities and allowing easy analysis of the dispersive and
absorptive responses of our model system to light and magnetic
field. Perturbative solutions for optical coherences can be
obtained in the low light power regime, $G\ll 1$, by expanding
relation (\ref{przed_rozkladem}) into power series of $G$.

Relation (\ref{przed_rozkladem}) contains the ratio of two
polynomials of $\delta$ (they also depend on $\omega_B'$). At this
point atomic movement needs to be taken into consideration, i.e.
$\delta$ has to be corrected for Doppler shifts $(\delta
\rightarrow \delta-kv)$ and expression (\ref{przed_rozkladem})
needs to be integrated over the Maxwell distribution of the
velocities $v$. Generally, this is the place when one sacrifices
analyticity of solutions for a single velocity class. However, the
benefits of analytic expressions can be retained by expanding
(\ref{przed_rozkladem}) into partial fractions
\begin{equation}\tilde\rho_{aB}=-n_{eq}\beta[\frac{A_+(\omega_B')}
{\delta+\delta_1(\omega_B')}+\frac{A_-(\omega_B')}{\delta-\delta_1(\omega_B')}+
\frac{B_+(\omega_B')}{\delta+\delta_2(\omega_B')}+\frac{B_-(\omega_B')}
{\delta-\delta_2(\omega_B')}\,],\label{koherencja_bez_ruchu}\reseteqn
\end{equation}
\noindent where $\delta_{1,2}^2=B^2-\gamma^2-rG\pm
\sqrt{r^2G^2-4B^2(\gamma^2+rG)-2rG^2}$ are the roots of the
denominator of Eq. (\ref{przed_rozkladem}), while $A_{\pm}$,
$B_{\pm}$ are the expansion coefficients equal to
$$
A_{\pm}=\frac{1} {2}\frac{\delta_1^3\mp a\delta_1^2+b\delta_1\mp
c} {\delta_1 (\delta_1^2-\delta_2^2)},\ \ \ B_{\pm}=-\frac{1}
{2}\frac{\delta_2^3\mp a\delta_2^2+b\delta_2\mp c} {\delta_2
(\delta_1^2-\delta_2^2)},$$ \noindent where $a=B+i\gamma$,
$b=2rG-(a^*)^2$, and $c=-a^*(2rG+|a|^2)$. One thus obtains
\reseteqn
\begin{eqnarray}\langle\tilde\rho_{aB}\rangle_v=-\frac{n_{eq}\beta N_0}{\sqrt{\pi}u}
[A_+(\omega_B')\int_{-\infty}^\infty\frac{\exp(-v^2/u^2)\;dv}{\delta-kv+\delta_1(\omega_B')}\nonumber\\
+A_-(\omega_B')\int_{-\infty}^\infty\frac{\exp(-v^2/u^2)\;dv}{\delta-kv-\delta_1(\omega_B')}\nonumber\\
+B_+(\omega_B')\int_{-\infty}^\infty\frac{\exp(-v^2/u^2)\;dv}{\delta-kv+\delta_2(\omega_B')}\nonumber\\
+B_-(\omega_B')\int_{-\infty}^\infty\frac{\exp(-v^2/u^2)\;dv}{\delta-kv-\delta_2(\omega_B')}\,],
\label{koherencja_po_predkosciach}\end{eqnarray} \noindent where
$u$ is the average atomic speed. Writing down the coherence in
form (\ref{koherencja_po_predkosciach}) allows a clear
interpretation of each contribution. The functions
$A_{\pm}(\omega_B')$, $B_{\pm}(\omega_B')$ are responsible for
narrow structures in the FS signals and describe the nonlinear
Faraday effect, while the integrals describe wide pedestals of
$I_{FS}(\omega_B')$, barely visible in a narrow scan of the
magnetic field. For calculations of the integrals in
(\ref{koherencja_po_predkosciach}), standard tabularized
functions, the plasma dispersion function \cite{Plasma} or the
complex error function \cite{Abramovitz}, can be used.

As it was mentioned above, the FS signal registered with the
crossed polarizers arrangement results from competition between
rotation of the polarization plane and net absorption. The
competition is seen in Eq.(\ref{FS}) which for $\delta=0$, takes
the form
\begin{equation}I_{FS}=I_0\sin^2(\frac{n_+-n_-}{2}kL)\exp(-2\kappa kL)
\label{natezenie},\reseteqn\end{equation} \noindent where
$\kappa=\kappa_+(\omega_B, \delta=0)=\kappa_-(\omega_B,
\delta=0)$. Relation (\ref{natezenie}) was used for simulation of
the FS signals with parameters corresponding to the experimental
conditions: $d$ = 2.25 $ \times 10^{-29}$ Cm, $u$ = 200 m/s, $g_e$
= $g_g$ = 1, $\gamma_g$ = 2$\pi\ \times\ 0.1\ $MHz, $\gamma_e$ =
2$\pi\times\ 5.74\ $MHz, and $\Gamma\ $= 2$\pi\ \times\ 2.92\
$MHz. The ground-state relaxation rate $\gamma_g$ reflects finite
atomic transit time across the laser beam. In the $D_1$ line of
rubidium, there are two transitions within the Doppler width that
can be excited by light: $F_g=3\rightarrow F_e=2$ and
$F_g=3\rightarrow F_e=3$ but since in our experiment the laser was
tuned only to the $F_g=3\rightarrow F_e=2$ component, the
contribution of another transition has been neglected. Apart from
a very simplified atomic structure, the present analysis also does
not take into account specific relaxation mechanisms, e.g.
velocity-changing collisions\cite{Arim_VCC} and/or radiation
trapping \cite{Matsko} which may affect the specific ratio
$\alpha/\beta$ in Eq. (\ref{alpha}).

Results of the simulations with our X model of the FS signal
amplitude vs. atomic concentration are shown in Fig. 6 for various
light intensities.

As the experimental results were recorded vs. vapor temperature,
the atomic density in Eq. (\ref{koherencja_po_predkosciach}) has
to be expressed as a function of the vapor temperature which is
possible if the temperature dependence of rubidium vapor pressure
is known. The issue of vapor pressure is crucial for many
experimental situations, yet exact measurements of vapor pressures
are difficult and subject to various systematic errors. A common
procedure is to use one of the existing compilations which fit
experimental data obtained in various temperatures by simple
analytical formulae that can be used for extra- or interpolation
to other temperatures. In principal, such fits are based on the
statistical physics and chemistry laws (Arrhenius law, ideal gas
law) with possible corrections for departures from the ideal gas
model and/or for specific metal evaporation conditions. Particular
care should be taken close to phase-transition temperatures, as
e.g. near the melting point. This is the case of our experimental
conditions as the melting point of rubidium $39.3\ ^o$C occurs
just in the middle of our temperature range.

Assuming ideal gas and starting from the Arrhenius relation, the
saturated vapor pressure could be determined by formula\reseteqn
\begin{equation}\log_{10}p(T)=A - \frac{B}{T}.\label{CRC}\end{equation}

Such a dependence is used, e.g. in reference\cite{CRC}, which for
rubidium gives $A=4.875$ and $B=-4215$ below the melting point and
$A=4.312$ and $B=-4040$ above the melting temperature. Another
widely cited compilation \cite{Nesmeyanov} applies extra
corrections terms and uses relation\reseteqn
\begin{equation}\log_{10}p(T)=A'+\frac{B'}{T}+C'T+D'\log_{10} T,\label{Nesm}\end{equation}
\noindent where for rubidium $A'=94.05$, $B'=1961$, $C'=-0.03772$,
and $D'=42.57$ below the melting temperature and $A'=15.88$,
$B'=4330$, $C'=0.00059$, and $D'=-2.99$ above the melting point.
From Eqs. (\ref{CRC}) and (\ref{Nesm}), with the help of ideal gas
equation, $pV=N_{at}k_BT$ ($N_{at}$ is the number of atoms the
gas, $V$ is its volume, and $k_B$ the Boltzmann constant), one can
obtain the final concentration-temperature relation $N_0(T)$.
Reliability of such density determination depends obviously on
quality of experimental data used for a given analytic fit.
Reference\cite{Koncen} presents a thorough comparison of several
measurements of rubidium vapor pressure vs. temperature which
appears to be fairly consistent with the Nesmeyanov's
prediction\cite{Nesmeyanov}. On the other hand, the CRC
compilation\cite{CRC} is based on not too many experiments and, to
our best knowledge, has not been verified in a wide range of
temperatures.

Our experiment interpreted in terms of the X model yields for
$T\approx 50\ ^o$C the density that is consistent with the data
given by\cite{CRC} and by 25\% higher from the value given in
reference\cite{Nesmeyanov}. We hope that after more detailed
interpretation, the described method could allow accurate
measurements of the atomic vapor densities and possibly improve
existing literature inconsistencies.

Comparison of Fig. 3 with Fig. 7 shows that while there is not a
perfect agreement of the simulated signals with the experimental
ones, they are very similar and exhibit the same general
character. For low temperatures the FS amplitude scales as a
square function of the concentration $N_0$. This part of the
dependences is determined only by the dispersive properties of the
medium. For higher temperatures the absorption starts to play an
important role and cannot be neglected anymore. The interplay
between dispersion and absorption produces the maximum of the FS
amplitude signal. For even higher temperatures, the absorption of
the transmitted light becomes dominant and the amplitudes of the
FS signals decrease exponentially with concentration $N_0$. In
addition to this temperature/density dependence, the FS amplitude
rises and shifts with increase of the incident light intensity.
The shifts of the maxima positions for various light intensities
are due to different saturation behavior of the dispersion and
absorption coefficients which affects the $\alpha/\beta$ ratio in
Eq. (\ref{alpha}). This is a very interesting effect that deserves
independent systematic study, preferably with more complex models
of atomic structure.

\section{Conclusions}
The specific effect of competition between polarization rotation
and absorption, described above, shows up in the FS signals. The
FS intensity, given by Eq. (\ref{natezenie}), increases with the
medium birefringence (as $\sin^2$ of the Faraday angle) and
decreases exponentially with the medium absorption.  Competition
of these opposite trends is responsible for the observed maxima of
the dependences seen in Figs. 3 and 7. Appearance of this
competition is an universal feature, independent of whether the
light-matter interaction has a linear or a nonlinear character.
The possible nonlinearity only affects the absolute scale of the
rotation and absorption and, consequently, the range of densities
where the competition occurs. For alkali metal vapors and for
light intensities of the order of 1 mW/cm$^2$, the resonant
nonlinear Faraday effect allows observation of such a competition
already in a low-density range, about $N_0\approx 10^{10}$
at/cm$^3$, which requires temperatures not much higher than the
room temperature. The onset of optical nonlinearity not only
lowers the density at which the characteristic maximum of the
$A_{FS}(N_0)$ dependence appears but results in its shift with the
light intensity. This is due to different saturation behavior of
the dispersive and absorptive atomic properties.

We note that the main features of the discussed dependences do not
depend on the detailed mechanism of the nonlinearity. We have
considered two physically different models in which the
nonlinearities are caused by two different mechanisms: either they
are due to redistribution of populations between Zeeman sublevels
(optical pumping), or are due to the coherence established between
the sublevels. Both models gave very similar results and
reproduced our experimental observations. The very distinct result
of the competition between absorption and rotation, i.e. the
maximum of the forward scattering amplitude signal, appears at
well determined value of the atomic density. We believe that
existence of such characteristic point whose position is well
defined function of the atomic properties of the medium and
scattered light intensity should allow precise determination of
the atomic number densities or transition probabilities.

\section{Acknowledgements}
We acknowledge stimulating and helpful discussions with Dmitry
Budker, Krzysztof Sacha and Jerzy Zachorowski whose experimental
help is also very appreciated. This work was partially supported
by NRC U.S.-Poland Twinning Grant No.015369, and by the Polish KBN
grant No. 3T11B 079 26. It was also a part of a program of Polish
National Laboratory of AMO Physics in Toru\'{n}, Poland (grant No.
PBZ/KBN/043/PO3/2001).

\bibliographystyle{amsplain}

\newpage

  \begin{figure}[h]\centerline{\scalebox{1}{\includegraphics{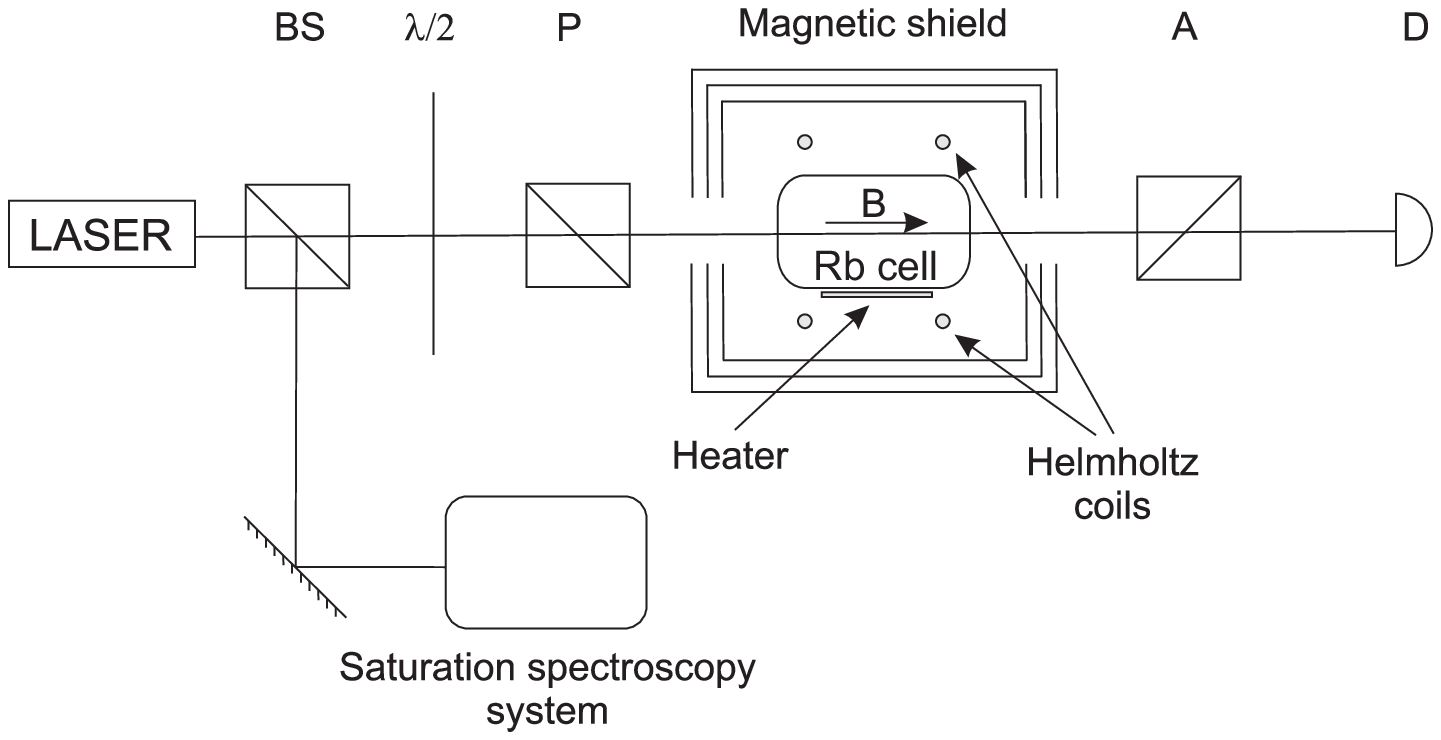}}}
  \caption{Experimental setup. BS is the beam-splitter, P, A are
  polarizer and analyzer, respectively, $\lambda/2$ is
$\lambda/2$-waveplate and D is a photodetector.}
  \end{figure}

\newpage

  \begin{figure}[h]\centerline{\scalebox{1}{\includegraphics{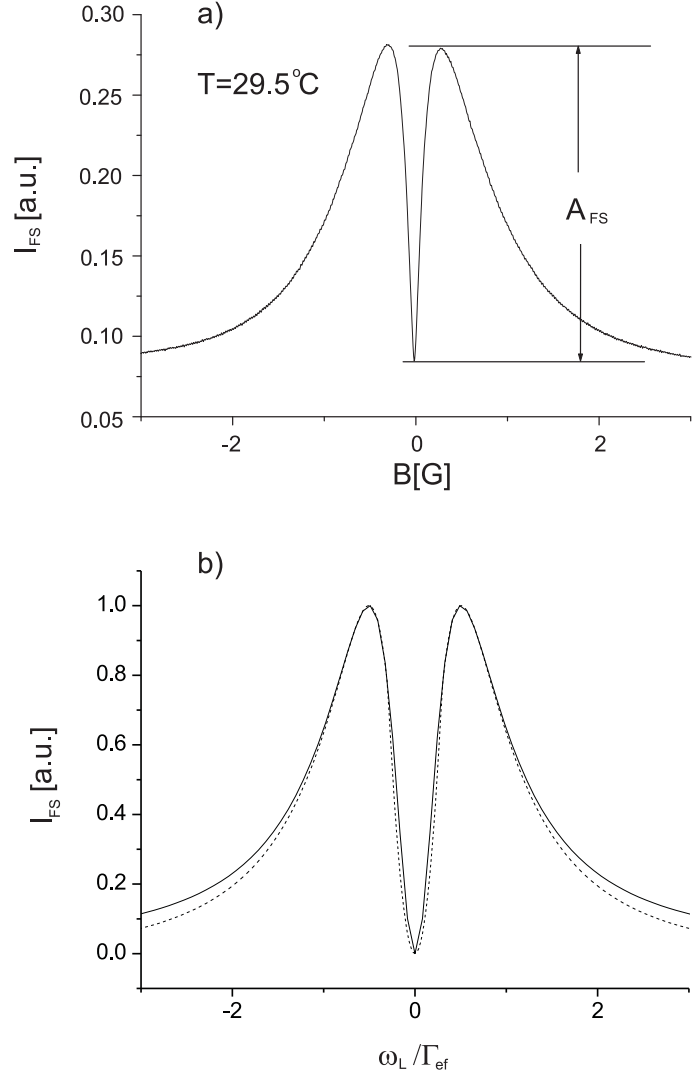}}}
  \caption{Typical forward scattering signal, $I_{FS}(B)$, recorded for
  cell temperature 29.5 $^o$C and light intensity $I_0$ = 2.2 mW/cm$^2$
  (a). FS signals calculated with the X model (solid line) and
  $\Lambda$ model (dashed line) (b). A$_{FS}$ denotes the amplitude of
  the FS signal. The theoretical signals correspond to incident light
  intensity $I_0=0.6$ mW/cm$^2$. They are presented in domain of Larmor frequency
  over effective width ($\Gamma_{ef}=\Gamma$ for the X model and
  $\Gamma_{ef}=\gamma_g$ for the $\Lambda$ model) and are normalized to the same
  $A_{FS}$.}
  \end{figure}

\newpage

  \begin{figure}[h]\centerline{\scalebox{1}{\includegraphics{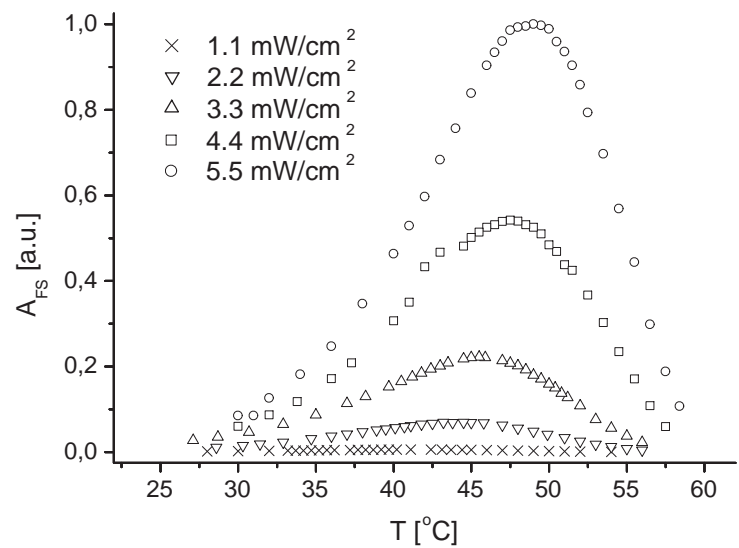}}}
  \caption{The amplitudes of the FS signals ($A_{FS}$) vs. temperature for
  various intensities of the incident light.}
  \end{figure}

\newpage

  \begin{figure}[h]\centerline{\scalebox{1}{\includegraphics{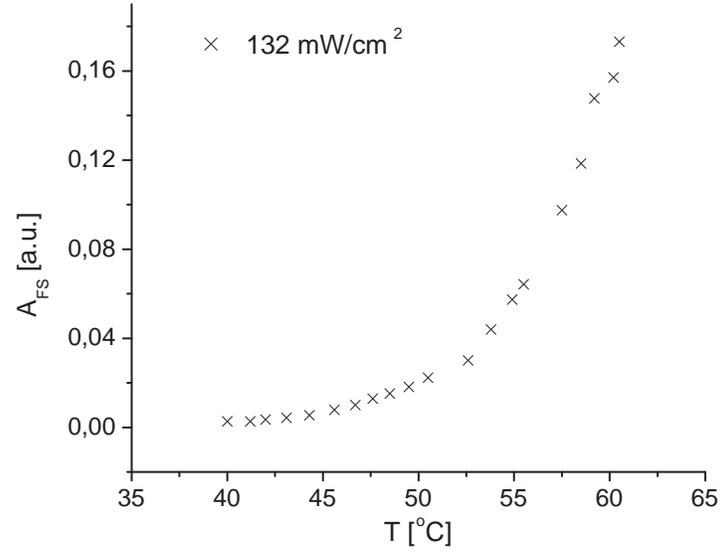}}}
  \caption{The amplitude of the FS signal (A$_{FS}$) vs. temperature for very
  high light intensity, $I_0$ = 130 mW/cm$^2$.}
  \end{figure}

\newpage

  \begin{figure}[h]\centerline{\scalebox{1}{\includegraphics{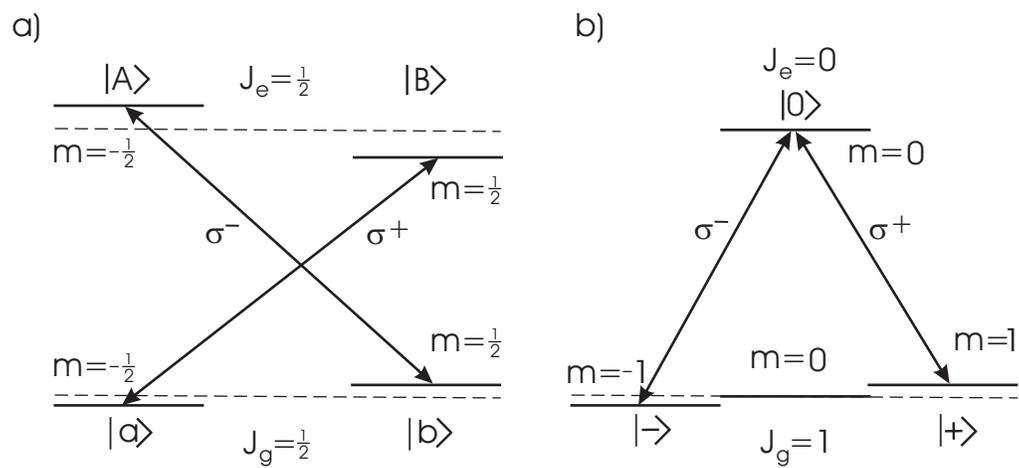}}}
  \caption{Energy level structure for a) the X-model, b) the $\Lambda$-model.}
  \end{figure}

\newpage

  \begin{figure}[h]\centerline{\scalebox{1}{\includegraphics{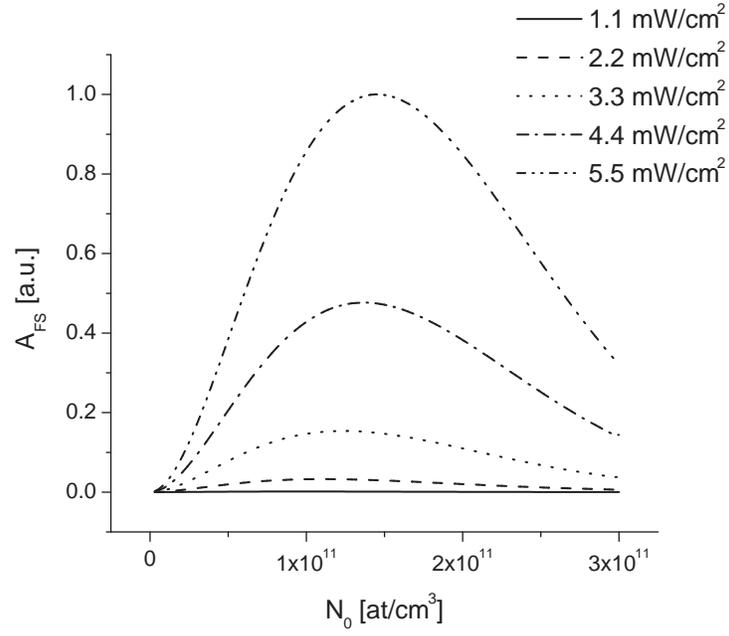}}}
  \caption{Results of numerical simulations of
  the FS signal amplitudes vs. atomic
  concentration for the same incident light intensities as used in experiment.}
  \end{figure}

\newpage

  \begin{figure}[h]\centerline{\scalebox{1}{\includegraphics{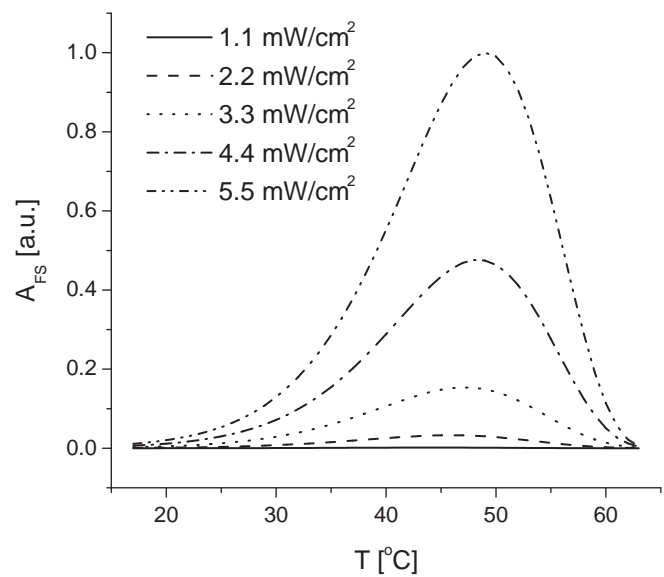}}}
  \caption{Results of numerical simulations of
  the FS signal amplitudes vs. vapor temperature for the light
  intensities as used in the experiment (compare with Fig. 3).}
  \end{figure}

\end{document}